# Android based Portable Hand Sign Recognition System

Jagdish L. Raheja,  A. Singhal*, Sadab*

Machine Vision Lab

CSIR-CEERI, Pilani

India

jagdish@ceeri.res.in, {asinghal, sadab1933}@gmail.com

Ankit Chaudhary

Dept. of Computer Science

Truman State University

USA

dr.ankit@ieee.org

**Abstract:**  These days mobile devices like phones or tablets are very common among people of all age. They are connected with network and provide seamless communications through internet or cellular services. These devices can be a big help for the people who are not able to communicatie properly and even in emergency conditions. A disabled person who is not able to speak or a person who speak a different language, these devices can be a boon for them as understanding, translating and speaking systems for these people. This chapter discusses a portable android based hand sign recognition system which can be used by disabled people. This chapter shows a part of on-going project. Computer Vision based techniques were used for image analysis and PCA was used after image tokenizer for recognition. This method was tested with webcam results to make system more robust.

**Keywords:** *Portable Gesture Recognition Systems, Android, Sign Recogntion, Edge Detection, Disabled People, Mobile Applications,*

*\* Student trainee at CSIR-CEERI.*

## Introduction

Image processing is a rapidly growing area in diverse applications, such as multimedia computing, secured data communication, biomedical, biometrics, remote sensing, texture understanding, pattern recognition, content-based retrieval, compression, and many more. This is all about how a computer can sense pictorial data after processing an image. Among the set of gestures intuitively performed by humans when communicating with each other, pointing gestures are especially interesting for communication and is perhaps the most intuitive interface for selection. They open up the possibility of intuitively indicating objects and locations, e.g., to make a robot change moving direction or simply mark some object. This is particularly useful in combination with speech recognition as pointing gestures can be used to specify parameters of location in verbal statements.

This technology can be a boon for disable people who are not able to speak hence cant communicate. Also if the person has different language than receiver, than also, it can be used to as translator. There has been always considered a challenge the development of a natural interaction interface, where people interact with technology as they are used to interact with the real world. A hand free interface, based only on human gestures, where no devices are attached to the user, will naturally immerse the user from the real world to the virtual environment.

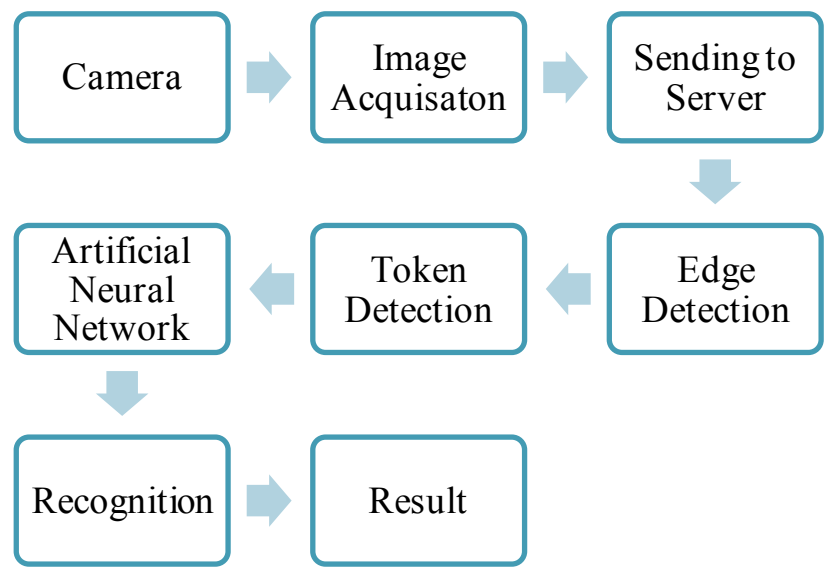

Figure 1: Block diagram of System

In this chapter, an application for the disable people has beed discussed which ahs been developed on Android device. Android device brings the long-expected technology to interact with graphical interfaces to the masses. Android device captures the user's movements without the need of a controller. The basic block diagram of a hand gesture recognition system is given in Figure 1. The video sequences captured by an android camera are processed to make it suitable for extracting useful information about user. In this work, a real time object selection and recognition method is proposed for the application for disabled people.

**Hand Gesture Recognition**
In the past decade, the computational power of computers has doubled, while the human computer interface (HCI) has not changed too much. When we work with a computer, we are constrained by intermediary devices (keyboards and mice). However, these are inconvenient and have become a bottleneck in human-computer interaction. In our daily life, we use speech to communicate with each other, and use gestures to point, emphasize and navigate. They are the more natural and preferable means to interact with computers for human beings. To make computers understand this however is not an easy task.

Gesture recognition is a topic in computer science and language technology with the goal of interpreting human gestures via mathematical algorithms. Gestures can originate from any bodily motion or state but commonly originate from the face or hand. Gesture recognition can be seen as a way for computers to begin to understand human body language, thus building a richer bridge between machines and humans. Gesture recognition enables humans to communicate with the machine and interact naturally without any mechanical devices. Gesture recognition can be conducted with techniques from computer vision and image processing.

Gestures of the Hand are read by an input sensing device such as an mobile or computer. It reads the movements of the human body and communicates with computer that uses these gestures as an input. These gestures are then interpreted using algorithm either based on statistical analysis or artificial intelligence techniques. The primary goal of gesture recognition research is to create a system which can identify specific human hand gestures and use them to convey information. By recognizing the hand symbols of a man it can help in communication with deaf and dumb people. It helps in taking prompt action at that time. Many Hand Gesture recognition projects are discussed in [1-5]. Many researchers [6-8] have tried with different instruments and equipment to measure hand movements like gloves, sensors or wires, but in these techniques user have to wear the device which doesn't make sense in practical use. So people thought about a way of contact less gesture recognition that could be considered as a research area in Computer Vision and which would be as natural as human to human interaction [9].

Gestures are usually understood as hand and body movement that can pass information from one person to another. Since we consider only hand gestures, the movement of the hand that expresses or emphasizes an idea, sentiment or attitude belongs to a gesture. Hand gestures can be classified into four categories according to the different application scenarios: conversational gestures, controlling gestures, manipulative gestures and communicative gestures [1]. We use conversational gestures to help express ourselves more clearly in our daily life. They are very subtle and need careful psychological studies. Controlling gestures can be designed to navigate in a virtual environment. For example, we can ask the computer to drive a car to the south by pointing in that direction. Manipulative gestures serves as a natural way to interact with virtual objects[10-12]. Sign language is an important case of communicative gesture [13]. Deaf people rely on it to talk to each other. It is objective and well defined and rarely causes ambiguity which makes it suitable as a test-bed for gesture recognition systems.

The approaches can be broadly classified into "Data-Glove based" and "Vision-based". Many recognition systems are based on the data-glove, an expensive wired electronic device. Various sensors are placed on the glove to detect the global position and relative configurations of the hand. One limitation of this method is the price of glove and other problem with this approach is that it needs a physical link between user and the computer. Because of this shortcoming, more and more researchers show interest in vision-based systems, which are wireless and the only thing needed is one or multiple cameras. In this work we present a vision-based system using one camera in android device. Digital image processing allows a much wider range of algorithms to be applied to the input data, and can avoid problems such as the build-up of noise and signal distortion during processing [14].

## Proposed Method

Signs are one form of hand gestures. Sign language is used as a communication medium among deaf & dumb people to convey the message with normal person. A person who can talk and hear properly (normal person) cannot communicate with deaf & dumb person unless he/she is familiar with sign language. Same case is applicable when a deaf & dumb person wants to communicate with a normal person or blind person. In order to bridge the gap in communication among deaf & dumb community and normal community, Video Relay Service (VRS) is being used nowadays. In VRS a manual interpreter translates the hand signs to voice and vice versa to help communication at both ends. A lot of research work has been carried out to automate the process of sign language interpretation with the help of image processing and pattern recognition techniques.

A lot of researchers initially used morphological operations to detect hand from image frames. It is an important research area not only from engineering point of view but also for its impact on the society. Sign languages are non verbal visual languages, different from spoken languages, but they serve the same function. There are different sign languages all over the world such as American Sign Language (ASL), British Sign Language (BSL), Japanese Sign Language family, French Sign Language family, Australian Sign Language, Indian Sign Language (ISL) etc.

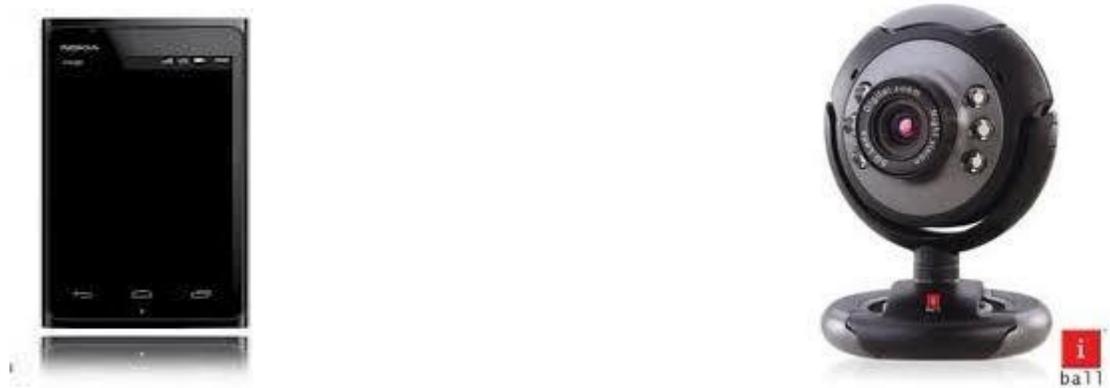

Figure 2: Android Device and iBall Camera

We are developing a system using an android device and also would be tested with webcam. This system will be useful for deaf and dumb person carrying an android device or a system with webcam. The Android device and iBall camera used in project are shown in Figure 2. The algorithms for detection using webcam and android are show in Figure 3 and Figure 4.

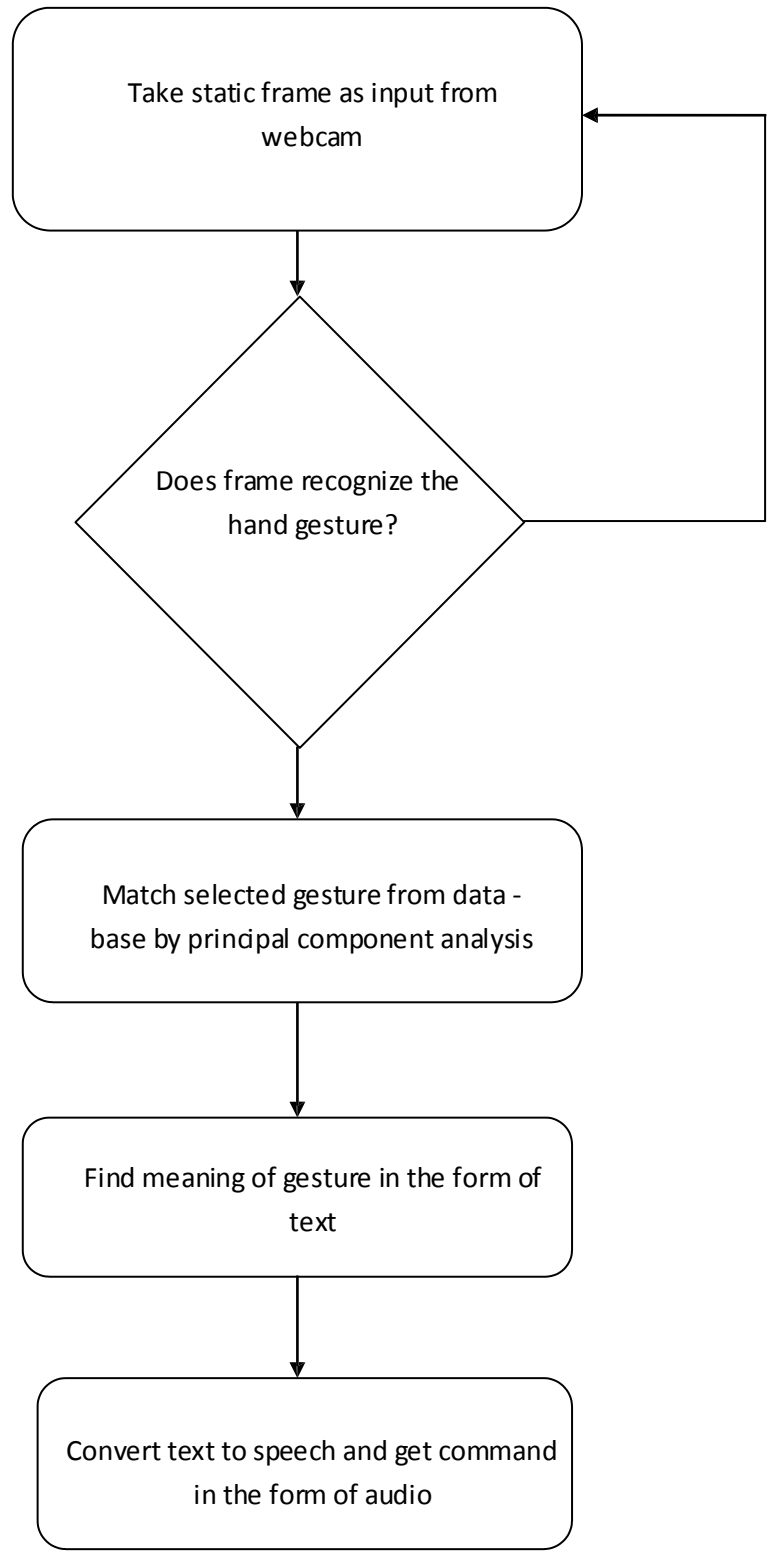

Figure 3: Block Diagram of Sign Language Recognition Model Using Webcam

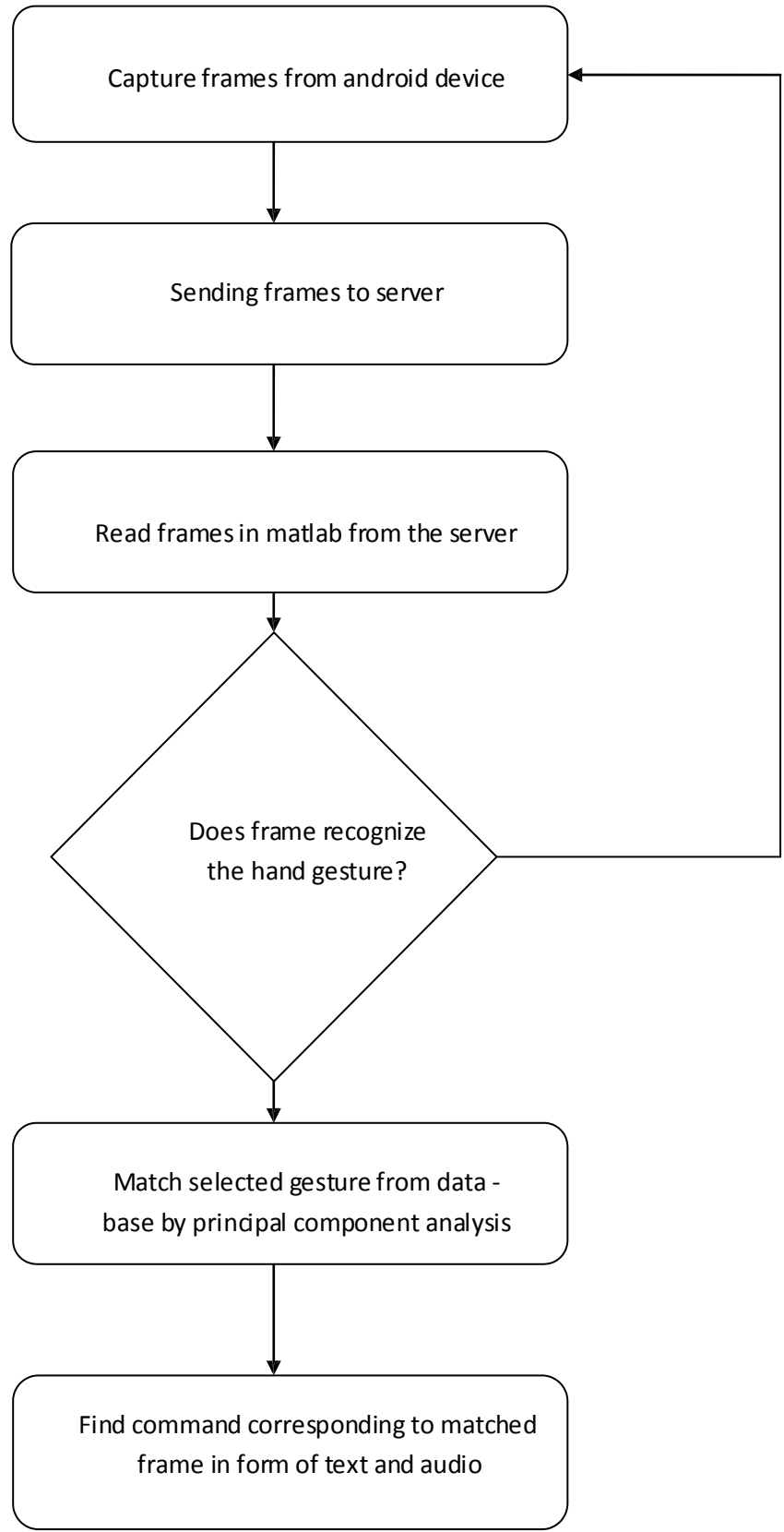

Figure 4 : Block Diagram of Sign Language Recognition Using Android Device

**Principal Component Analysis (PCA)**

Like face recognition, which is inherently a classification problem in a high dimensional feature space, we also treat the recognition of hand gestures as a problem in the field of pattern recognition, and indeed, many techniques have been proposed in this area. PCA [15] is a standard tool in modern data analysis - in diverse fields from neuroscience to computer graphics - because it is a simple, non-parametric method for extracting relevant information from confusing data sets. It is a way of identifying patterns in data, and expressing the data in such a way as to highlight their similarities and differences. Since patterns in data can be hard to find in data of high dimension, where the luxury of graphical representation is not available, PCA is a powerful tool for analyzing data [15]. The other main advantage of PCA is that once you have found these patterns in the data, and you compress the data, i.e. by reducing the number of dimensions, without much loss of information. This technique used in image compression. PCA is a rather general statistical technique that can be used to reduce the dimensionality of the feature space [16].

Given a set of training objects represented by their feature vectors, $x_i$, $y_i (1<i<n)$, where n is the number of samples, the training set can be written as

$$- \begin{bmatrix} x1, y1 \\ x2, y2 \\ x3, y2 \\ .... \\ xn, yn \end{bmatrix}$$

whose mean vector would be $x_i$ - mean(x) and $y_i$ - mean(y) and covariance matrix is $R_x$, defined as:

$$C^{n \times n} = (c_{i,j},\ c_{i,j} = cov(Dim_i, Dim_j)),$$

where $C^{nxn}$ is a matrix with n rows and n columns, and $Dim_x$ is the $x^{th}$ dimension.

The covariance matrix is a real symmetric square matrix of size N by N, where N is the length of the feature vector. The training set x,y corresponds to a cluster of data points in an N dimensional feature space. There exists redundancy in the feature space since the features, i.e the dimensions, are not independent of each other. By PCA we can eliminate the redundancy by transforming the original feature space into a so-called PC space in terms of Principal Components (PCs). Given the mean vector $\mathbf{x_m}$ and the covariance matrix $\mathbf{R_x}$ of a set of training samples, the eigen vector and eigenvalue would be calculated [15].

**Extracting Frames from Live Video**

This work was done for live realtime constraints, so we have to segment video into frames in such a memory efficient and fast way that complex processing on that frame can be done in real time for hand gesture detection. Here system capture frame per 1/3 second and three continuous frames are analyzed to determine motion in frame. It tries to capture the frame which is static. The difference between three continuous frames is less than specified value. Here we can say that hand gesture is shown in this frame.

For detecting frame of interest, following processing is applied: Suppose three frames are A, B, and C. Motion parameter = (A *XOR* B) *OR* (A *XOR* C) Suppose size of frame is M *X* N pixels.

If motion parameter is less than (M*N)/100 then frame A is considered as static and this frame is captured. This system is developed in MATLAB and in Java. It is tested and used successfully as a sub function in many work.

**Image Acquisition**

Image acquisition is the first step in any vision system. In this application it is done by using IPWebCam android application. The application uses the camera present in the phone for continuous image capturing and a simultaneous display on the screen. The image captured by the application is streamed over its Wi-Fi connection (or WLAN without internet as used here) for remote viewing.

The program access the image by logging to the device's IP, which is then showed in the GUI. The result of image capture is shown is Figure 5.

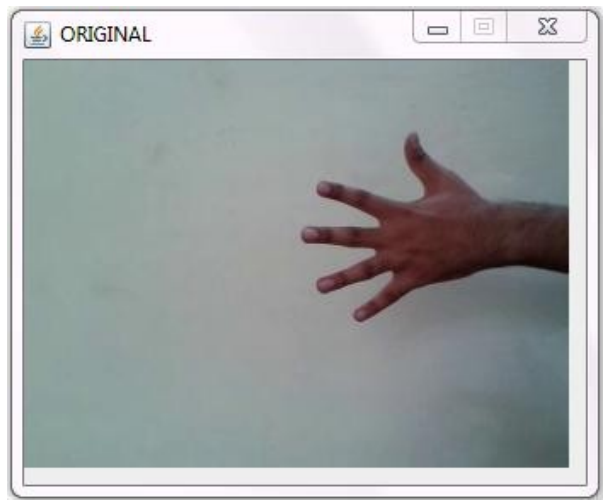

Figure 5: Original image captured from android device

**Edge Detection**

In this program the edge detection technique used is sobel edge detector, the mathematical working this has been explained in [17]. The image captured is then passed through sobel filter the results of that are shown in figure 6.

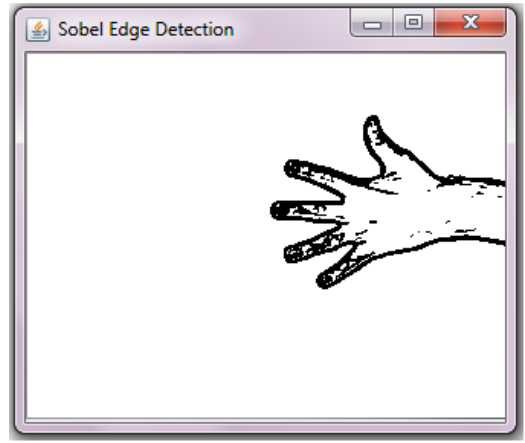

Figure 6: Sobel Edge Filtered Image

**Hand Token**

The idea here is to make the image into a neuronal network usable form is, that the cosinus and sinus angles of the shape represents the criteria's of a recognition pattern. Figure 7 show image with tokens. Each square represents a point on the shape of the hand image from which a line to the next square is drawn.

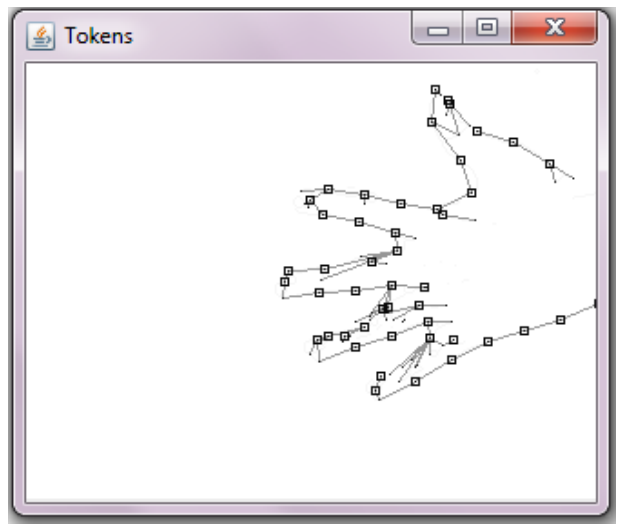

Figure 7: Generated token of the original image

On zooming a part of figure 7 it shows a right-angled triangle between the 2 consecutive squares, as shown in figure 8a. This and the summary of all triangles of a hand image are the representation of the tokens of a hand from which we can start the neuronal network calculations.

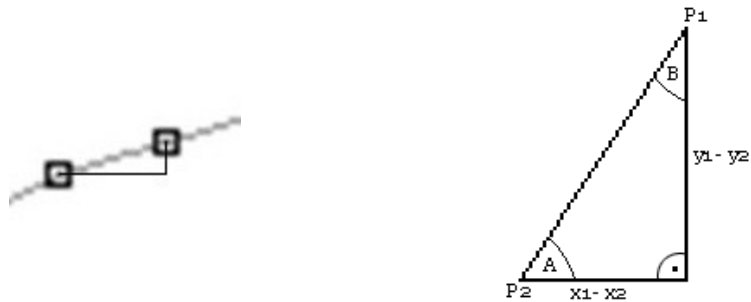

Figure 81: a. Zoomed image of image tokens & b. effective right angled triangle

The right-angled triangle in figure 8b represents a token of a single hand image. The angles A and B are the two necessary parts which will be fit into the neuronal network layers. With the two angles we can exactly represent the direction of the hypotenuse from point P1 to P2 which is represents the direction of a hand image.

 **Training Data**

Another main part of this work is the integration of a feed-forward backpropagation neural network. As described earlier the inputs for this neuronal network are the individual tokens of a hand image, and as a token normally consists of a cosinus and sinus angle, the amount of input layers for this network are the amount of tokens multiplied by two. The implemented network just has one input, hidden and output layer to simplify and speed-up the calculations on that java implementation. For training purpose the database of images located on the disk is used. It contains different types of pre-defined gestures. These gestures are shown in Figure 9 and also in Figure 11.  These gestures are first processed and then the tokens generated are passed to the network for training purpose. This process of training network from set images is done automatically when the application is initialized.

## Webcam Implementation

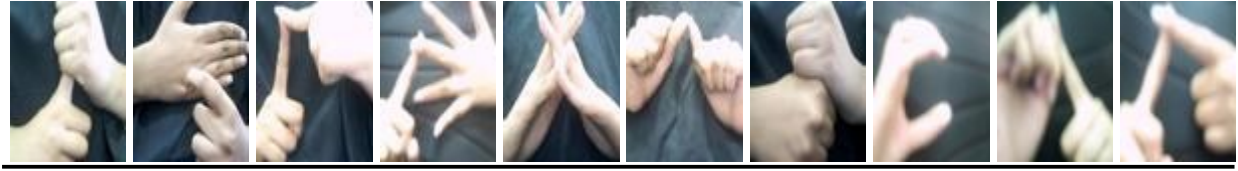

Figure 9: Sign Database

S R T H X A G C I E

Figure 10: Sign Database means

Database has been developed for 10 sign gestures using webcam as shown in Figure 9. These gestures have been taken from Indian Sign Languages that uses both hands except few gestures. Database of corresponding Alphabets has also been developed as shown in Figure 10. Gestures have been captured in black background with single light source. One can capture and store more than one image of gestures in database to improve the matching rate. Images in database should be resized into same dimension (60*80).

## Android Implementaion

A differetn database has been developed for 10 sign gestures using android device as shown in Figure 11. These gestures have been taken from Indian Sign Languages that uses both hands except few gestures. Database of corresponding Alphabets has also been developed are shown in Figure 12. Gestures have been captured in black background with single light source. One can capture and store more than one image of gestures in database to improve the matching rate. Images in database should be resized into same dimension (100*100).

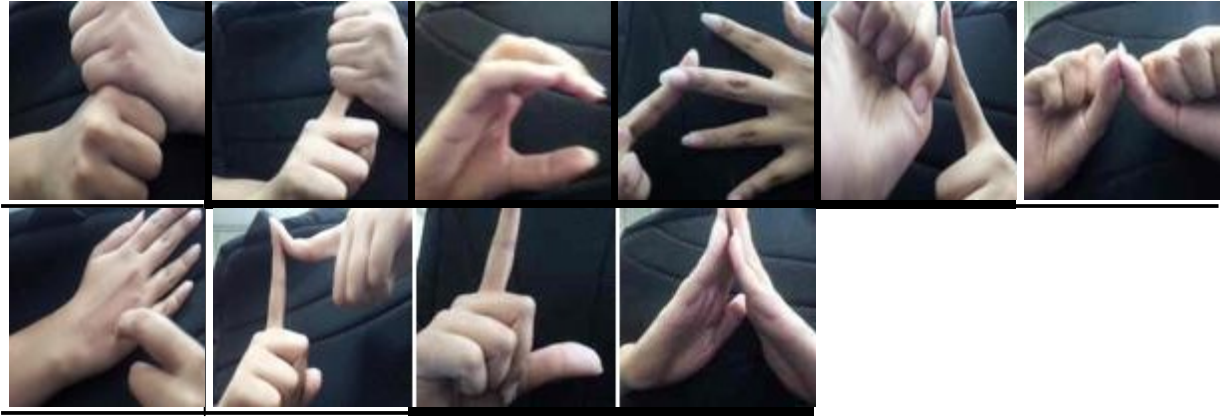

Figure 11: Sign Database

G S C H I A

R T L X

Figure 12: Sign Database means

**Recognition**

Recognition is the final step of the application. PCA was used to for this. Principal component was calculated and stored for database images. As any new gesture would be shonw to screen, its feature vector would be calculated and compred to database vectors. It gives percentage of recognition to each gesture with highest percentage closely matching and lowest to the farthest matching and the closest match is considered as the result. The match gesture would lead to actuation of the system. The recogntion system is shown in Figure 13.

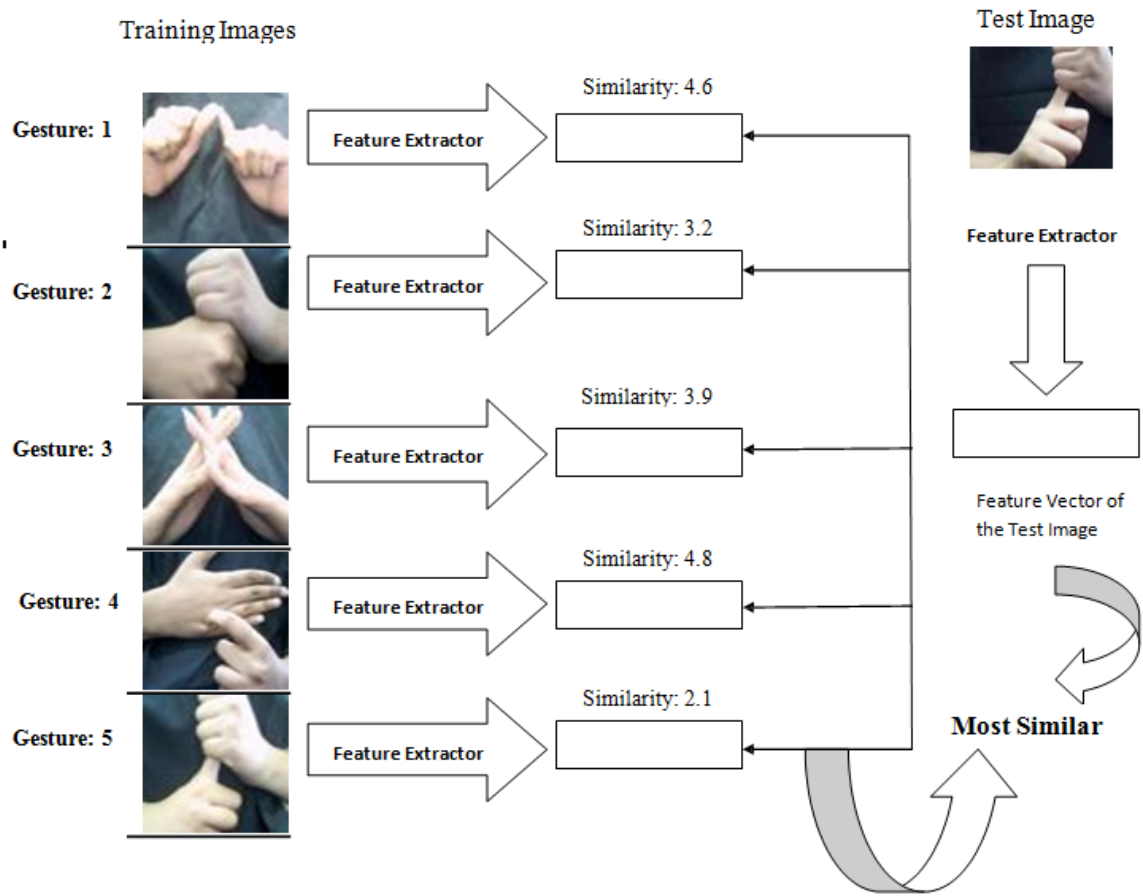

Figure 23: Recognition Process

**Results**

all input images were captured by webcam or an android device. The captured image is then read in matlab and converted in binary form of size defined so that it takes less time and memory space during pattern recognition. Now by using PCA it calculates the Eigen vectors and shows the equivalent image to the input gesture with minimum equivalent distance. At last it gives out the matched hand gesture and alphabet corresponding to the given input image from the database. It also produces an an audio sound indicating the output sign gesture.

The results of webcam based recogniton and android device systems are shown in Figure 14 and Figure 15 respectively. The accuracy for both methods are shwon in Table 1 and Table 2.

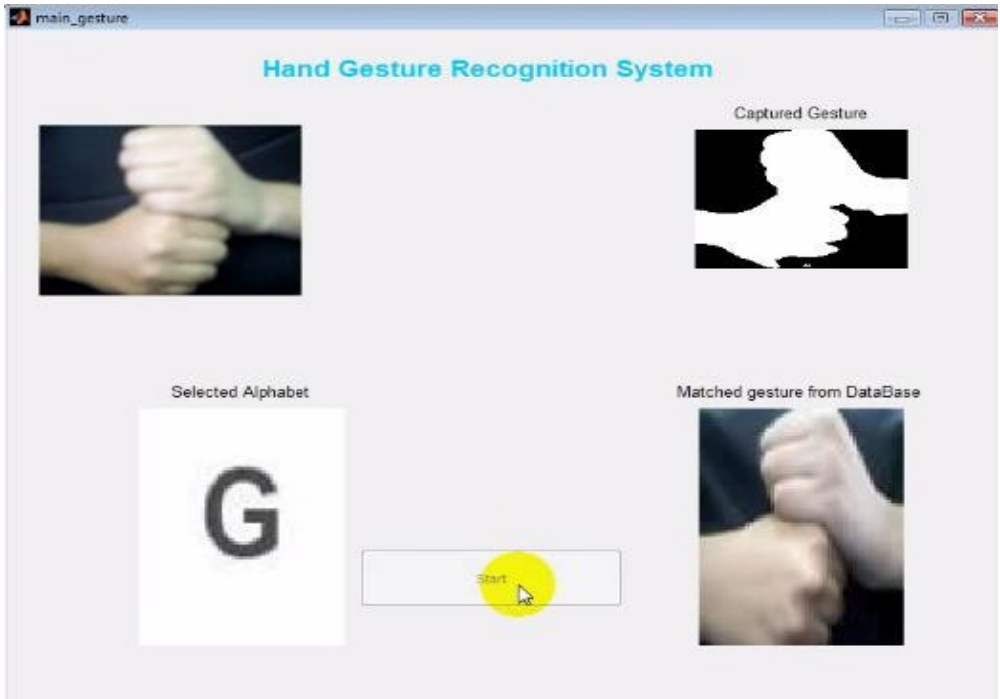
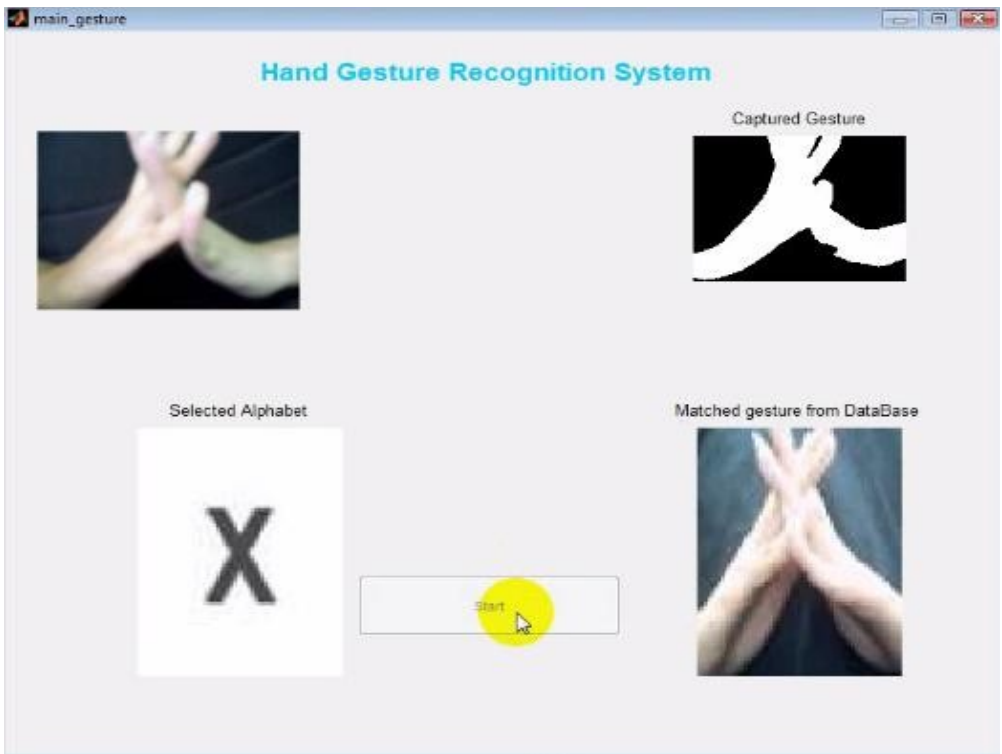

Figure 14: Test results with Webcam

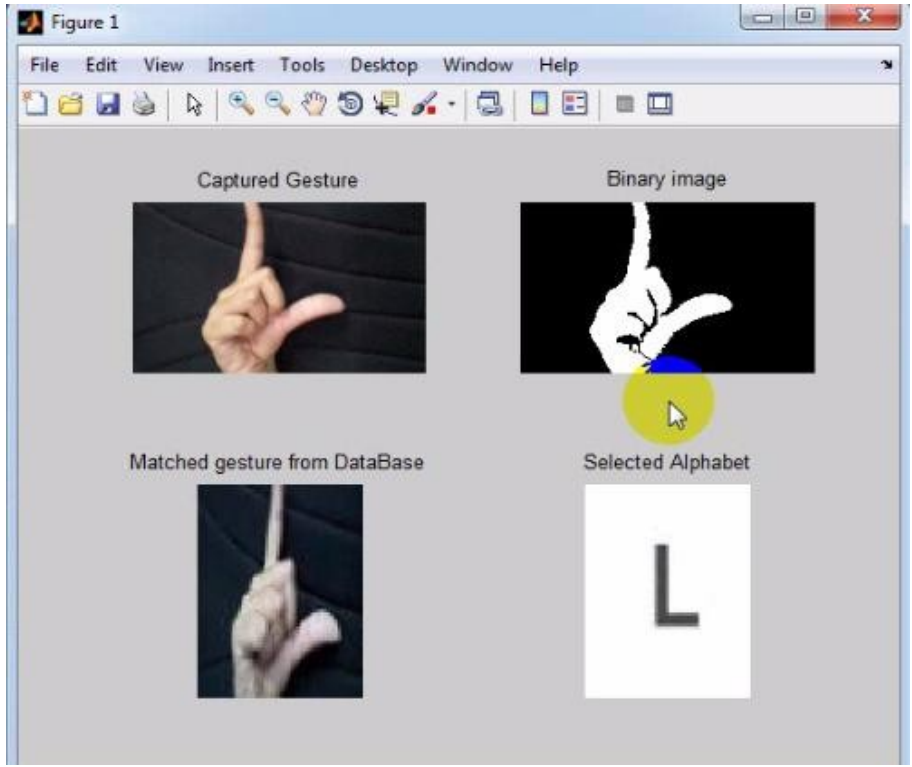
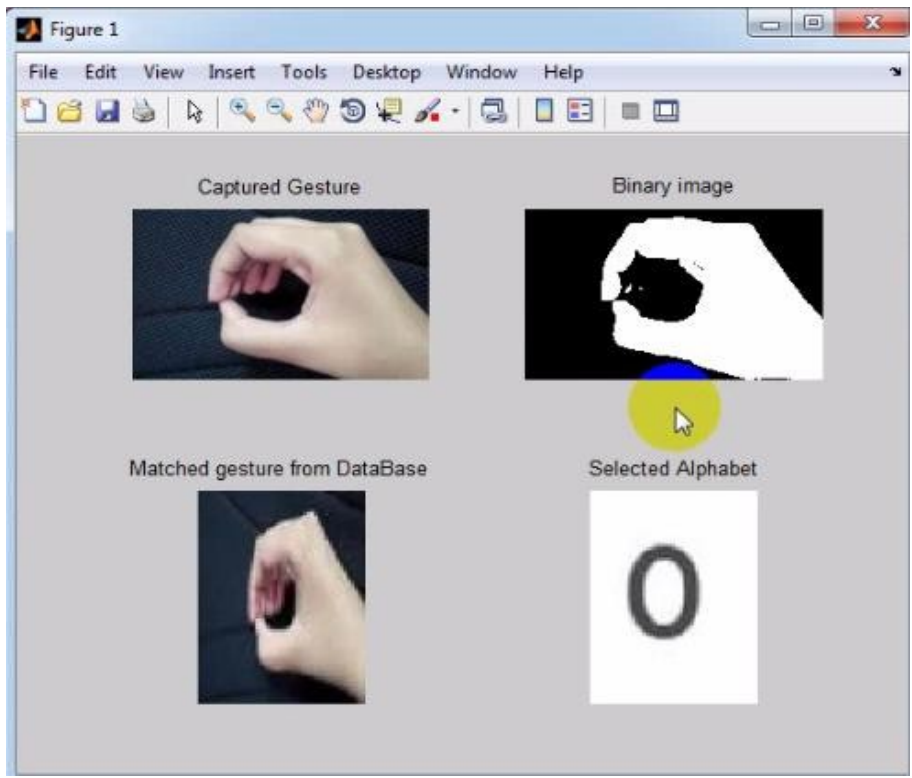

Figure 15: Test results with Android Device

Table 1: Test results with Webcam

| Gesture | Recognition Rate for 10 trials using webcam |
|---|---|
| A | 90% |
| C | 80% |
| E | 70% |
| I | 80% |
| G | 100% |
| H | 80% |
| R | 90% |
| S | 90% |
| T | 100% |
| X | 100% |

Table 2: Test results with Android Device

| Gesture | Recognition Rate for 10 trials using android device |
|---|---|
| A | 80% |
| C | 70% |
| G | 100% |
| H | 80% |
| I | 70% |
| L | 90% |
| R | 100% |
| S | 90% |
| T | 100% |
| X | 90% |

**Conclusion**

This chapter discuss a hand sign recogntion system which would be deployed on an Android device. The system is developed and tested successfully with webcam and an android device. This system is useful for a deaf and dumb person carrying an android device or a system connected with webcam. All gestures have recognition rate in between 70-100% which is an acceptable range. Overall accuracy of this system is 90% (approx) while 77% with Android.

The performance of the algorithm used for detecting sign gestures can be severely decreased due to varying lighting conditions and noises in the background [20-21]. In future, a custom camera instead of the IpWebCam app which will further enhance the success rate of the system. Other different type of gestures can also be made part of the database.